\documentclass[11pt]{article}
\usepackage[russian,english]{babel}
\usepackage{amssymb,amsthm}
\usepackage{amsmath}
\usepackage{inputenc}
\usepackage{graphicx}

\newcommand{\bc}{\begin{center}}
\newcommand{\ec}{\end{center}}
\newcommand{\be}{\begin{equation}}
\newcommand{\ee}{\end{equation}}
\newcommand{\bea}{\begin{eqnarray}}
\newcommand{\eea}{\end{eqnarray}}
\newcommand{\ba}{\begin{array}}
\newcommand{\ea}{\end{array}}
\newcommand{\edc}{\end{document}}

\begin{document}

\begin{center}
{\bf \large {ON GROUND STATES OF ROZIKOV MODEL ON THE CAYLEY TREE}}
\end{center}
\begin{center}
 G. I. BOTIROV
\end{center}

\textbf{\emph{ Abstract.}} In this paper we consider a model on a
Cayley tree which has a finite radius of interactions, the model was
first considered by Rozikov. We describe a set of periodic ground
states of the model.

\textbf{\emph{ The Cayley tree.}}

The Cayley tree $\Im^k$ of order $ k\geq 1 $ is an infinite tree,
i.e., a graph without cycles, such that each vertex of which lies on
$k+1$ edges. Let $\Im^k=(V,L,i)$, where $V$ is the set of vertexes
of $\Im^k$, $L$ is the set of edges of $\Im^k$, and $i$ is the
incidence function associating to each edge $l\in L$ its endpoints
$x, y \in V$. If $i (l) = \{ x, y \} $, then $x$ and $y$ are called
{\it nearest neighboring vertexes}, and we write $<x, y> $. A
collection of the pairs $<x_0, x_1>,<x_1, x_2>, \dots, <x_{d-1}, y>$
is called {\it{a path}} from $x$ to $y$. The distance $d(x,y), x,y
\in V$ is the length of the shortest path from $x$ to $y$ in $V$.

For the fixed $x^0 \in V$ we set $ W_n = \ \{x\in V\ \ | \ \ d (x,
x^0) =n \}, $ $$ V_n = \ \{x\in V\ \ | \ \ d (x, x^0) \leq n \},\
\ L_n = \ \{l = <x, y> \in L \ \ | \ \ x, y \in V_n \}. $$

It is known (see e.g. [2]) that there exists a one-to-one
correspondence between the set $V$ of vertices of the Cayley tree of
oreder $k\geq 1 $ and the group $G _ {k} $, of the free products of
$k+1$ cyclic groups $\{e, a_i\}, \ i=1, \dots, k+1$ of the second
order (i.e. $a^2_i=e, a^{-1}_i=a_i$) with generators $a_1, a_2,
\dots, a_{k+1}$.

\textbf{ \emph{ Configuration Space and the model}}

We consider models where the spin takes values in the set
$\Phi=\{1, 2, \dots , q \}, q \geq 2$. For $A \subseteq V$ a spin
configuration $\sigma_A$ on $A$ is defined as a function $x \in A
\rightarrow \sigma_A(x) \in \Phi$; the set of all configurations
coincides with $\Omega =\Phi ^ {V} $. We denote $\Omega=\Omega_V$
and $\sigma=\sigma_V$. Also we define a {\it{periodic
configuration}} as a configuration $\sigma \in \Omega$ which is
invariant under a subgroup of shifts $F_k \subset G_k$ of finite
index.

More precisely, a configuration $\sigma \in V$ is called $F_k$--
periodic if $\sigma(yx)=\sigma(x)$ for any $x \in G_k$ and $y \in
F_k$.

For a given periodic configuration the index of the subgroup is
called the period of the configuration. A configuration that is
invariant with respect to all shifts is called {\it
translational--invariant}.

For $A \subset V$ let us define a generalized Kronecker symbol (see
[6]) as the function $U(\sigma_A): \Omega_A \rightarrow \{|A|-1,
|A|-2, \dots , |A|-min\{|A|,|\Phi|\}\},$ by
$$ U(\sigma_A)=|A|-|\sigma_A \cap
\Phi|,\eqno (1)$$ where as before $\Phi =\{1,2, \dots, q\}$ and
$|\sigma_A \cap \Phi|$ is the number of different values of
$\sigma_A(x), x \in A$. For instance if $\sigma_A$ is a constant
configuration then $|\sigma_A \cap \Phi |=1$.

Note that if $|A|=2$, say, $A=\{x,y\}$, then
$U(\{\sigma(x),\sigma(y)\})=\delta_{\sigma(x)\sigma(y)}$,
$$\delta_{\sigma(x) \sigma(y)}=\left\{\begin{array}{ll} 1, \ \
 \sigma(x)=\sigma(y),\\
0,\ \ \sigma(x) \ne \sigma(y).\\
\end{array}\right.
$$

Fix $r \in N$ and put $r'=[\frac{r+1}{2}]$, where $[a]$ is the
integer part of $a$. Denote by $M_r$ the set of all balls
$b_r(x)=\{y \in V : d(x,y) \leq r'\}$ with radius $r'$, i.e.
$M_r=\{b_r(x):\ x \in V\}$.

We consider the energy of the configuration $\sigma \in \Omega$ is
given by the formal Hamiltonian $$H(\sigma)=-J \sum \limits_{b \in
M_r} U (\sigma_b), \eqno (2)$$ where $J \in R $. This Hamiltonian
was first considered by Rozikov [6].

\textbf{ \emph{ Ground states}}

The ground states for the model defined on $Z^d$ can, for example,
be found in [3], [7].

\textbf{ Definition 1.} A configuration $\varphi$ is called the
ground states of relative Hamiltonian $H$, if
$$U(\varphi_b)=U^{min}=min\{U(\sigma_b):\sigma_b
\in \Omega_b \} \  {\rm for \  any} \ b \in M_r.$$

In [1], [5] the ground states of Ising and Potts models with
competing interactions of radius $r=2$ on the Cayley tree were
described.

Let $GS(H)$ be the set of all ground states, and let $GS_p(H)$ be
the set of all periodic ground states.

\textbf{Theorem 1.} a) If $J>0$, then for all $r \geq 1$ and $k \geq
2$ the set $GS(H) $ consists only configurations $\{\sigma^{(i)},
i=1,2, \dots , s\}$, where $\sigma^{(i)} \equiv i, \forall x \in V$;

b) Let $r=2$, $J <0, \ \ q \geq 2^m $ and $k \in \{2 ^ {m-1}-1,
\dots , q-2 \}, \ m=3,4, \dots $ then there exists a normal subgroup
$F$ of index $2^m $, such, that any $F $ -- periodic configuration $
\sigma $ is a ground state for Hamiltonian $H$ i.e. $\sigma \in GS_p
(H) $.

\textbf {Proof} a) Easily follows from (1), (2) and Definition 1.

b) Since $J <0$ to construct a ground state it is necessary to
consider configurations $\sigma $ with a condition, that $U
(\sigma_b) =0$ for all $b \in M $, i.e. on any ball $b \in M $ the
configuration $\sigma$ is such that $\sigma(x)\ne \sigma(y)$ if
$x\ne y$. Therefore we will construct a normal subgroup $F$ of index
$2^m $ such, that any element of the set $S_1 (e) = \{e, a_1, \dots,
a _ {k+1} \} $ is not equivalent (with respect to $F$) to each other
element of the set. Since $k+2 \leq q $ we get $k \leq q-2$.
Consider a normal subgroup $F $ of index $2^m $, such that $F=F _
{A_1} \cap \dots \cap F _ {A_m} $ where $F _ {A_i} = \{x \in G_k:
\sum \limits _ {j \in A_i} \omega_j (x)-\mbox{even} \},$ and
$\omega_x (a_i)$ is the number of letter $a_i$, in nondeductible
word $x$, $ A_i \subset \{1, \dots, k+1 \}, i=1, \dots, m. $ Now we
shall construct $A_i, \ \ i=1, \dots, m$, so that all elements of
any ball $b \in M $ were from different classes of equivalency.

Let's consider all possible configurations $ \alpha: \{1,2, \dots, m
\} \rightarrow \{\mbox {e, o} \} $ (where "e" \ designates "even"
and "o" designates "odd"). Let's notice, that number of such
configurations is equal to $2^m $. From them choose half, i.e. $2 ^
{m-1} $ configurations with following properties: or the number of
letters \ "e" \ in a configuration is more than number of letters \
"o", or the number of letters \ "e" \ in a configuration is equal to
number of letters \ "o" \ and among the last there are no
configurations coinciding at replacement \ "e" \ on letters \ "o".
Let's denote these $2 ^ {m-1} $ configurations by
$$\alpha_0=\{ \mbox{e, \ e, \ e,  \dots , e}\}=(\alpha_{01}, \alpha_{02}, \dots , \alpha_{0m}) $$
$$\alpha_1=\{ \mbox{o, \ e, \ e, \dots , e} \}=(\alpha_{11}, \alpha_{12}, \dots , \alpha_{1m})$$
$$\alpha_2=\{ \mbox{e, \ o, \ e, \dots , e} \}=(\alpha_{21}, \alpha_{22}, \dots , \alpha_{2m})$$
$$\alpha_3=\{ \mbox{e, \ e, \ o, \dots , e} \}=(\alpha_{31}, \alpha_{32}, \dots , \alpha_{3m})$$
$$\dots \  \dots$$
$$\alpha_{2^{m-1}}=\{ \mbox{o, \ e, \ e, \dots , o}\}=(\alpha_{2^{m-1}1},\alpha_{2^{m-1}2},
\dots , \alpha_{2^{m-1}m}).$$

We can define sets $A_i, \ i=1,2, \dots, m $,  as follows

$$A_i=\{j \in \{1,2, \dots , k\}: \ \alpha_{ji} -
\mbox{odd}\} \cup \{k+1\}, \ i=1,2, \dots , m. \eqno(3)$$

Let's notice, that $A_i, \ i=1,2, \dots m $, make sense if $k+1 \geq
2 ^ {m-1} $ i.e. $k \geq 2 ^ {m-1}-1$. Check, that $F=F _ {A_1} \cap
\dots \cap F _ {A_m} $, constructed by sets (3), satisfies
conditions of the theorem. At first we shall prove, that $S_1 (e) $
with respect to $F$ divides into different non-equivalent elements:
Denote $ S_1 (x) = \{y \in V: d (x, y) =1 \} = \{x, xa_1, \dots, xa
_ {k+1} \}, \gamma_i(x) = |S_1 (x) \cap F_i | $. It is enough to
prove, that $ \gamma_i(x) =0 \ \mbox {or} \ 1$ for any $x \in V $
and $i=1, \dots, m $. By our construction one has $ \gamma_i (e) \in
\{0,1 \} $ for any $i = 1, \dots, m $. Hence, elements of the set
$S_1 (e) $ are not equivalent to each others, also they are not
equivalent to $e $. Then by Theorem 3 of [4] elements of the set
$S_1 (x)$ are not equivalent to each others. By Theorem 1 of [4] we
get $x \sim xa_i $ (i.e. $x $ and $xa_i $ belong to one class) if
and only if $e \sim a_i $. By our construction $e \nsim a_i, \forall
i=1, \dots, k+1$ hence $x \nsim xa_i $; therefore, $ \gamma_i (x) =0
\ \mbox {or} \ 1$.

The theorem is proved.

\textbf{Theorem 2.} Let $r=2.$ a) if $J>0$, then $|GS_p(H)|=q$;

b) If $J<0$, then $|GS_p(H)|=C^{k+2}_q(k+2)!$

\textbf{Proof.} Case a) is trivial. In case b) for a given
configuration $\varphi_b$, for which the energy $U(\varphi_b)$ is
minimal, we can use Theorem 1 to construct the periodic
configurations $\sigma$ with period $2^m$. In each case, the exact
number of such ground states coincides with the number of different
configurations $\sigma_b$, such that the energy  $U(\sigma_b)$ is
minimal for any $b \in M$. The theorem is proved.

 \vskip 0.3 truecm

{\bf Acknowledgements.} A part of this work was done at the ICTP,
Trieste, Italy and the author thanks ICTP for providing finicial
support and all facilities (July 2008).
 \vskip 0.3
truecm
\begin{center}
\textbf{References}
\end{center}

\begin{enumerate}\item{Botirov G.I., Rozikov U.A. Potts model with competing
interactions on the Cayley tree: the contour method
// \emph{Theor. Math. Phys.,} (2007), \textbf{153}, No 1, p. 1423-1433.}
\item{Ganikhodjaev N.N. \emph{Dokl. Akad. Nauk Resp. Uzbekistan}. (1994), \textbf{5}, No 4,
p. 3-5.}
\item{Minlos R.A., Introduction to Mathematical Statistical
Physics // \emph{Univ. Lecture Ser.,} (2000) \textbf{19}, AMS,
Providece, RI, ISSN 1047-3998.}
\item{Rozikov U.A. Partition structures of the Cayley tree and applications for describing
periodic Gibbs distributions // \emph{Theor. Math. Phys.,} (1997),
\textbf{113}, No 1, p. 929-933}
 \item{ Rozikov U.A. Constructive description
of ground states and Gibbs measures for Ising  model with two-step
interactions on Cayley tree // \emph{J. Stat. Phys.,} (2006),
\textbf{122}, No.2, p. 217-235.}
\item{ Rozikov U.A. A contour
method on Cayley tree
// \emph{J. Stat. Phys.,} (2008), \textbf{130}, p. 801-813.}
\item{Sinai Ya.G. Theory of phase transitions: Rigorous Results // Pergamon, Oxford, (1982)}

\end{enumerate}

\end{document}